# Standards-based World-Wide Semantic Interoperability for IoT

Ernö Kovacs[1], Martin Bauer[1], Jaeho Kim[2,3], Jaeseok Yun[2], Franck Le Gall[4], Mengxuan Zhao[4]

[1] NEC Laboratories Europe, Heidelberg, Germany,
[2] Korea Electronics Technology Institute, Gyeonggi-do, Korea,
[3] Yonsei University, Seoul, Korea,
[4] Easy Global Market, France

Email: {ernoe.kovacs, martin.bauer}@neclab.eu, {jhkim, jaeseok}@keti.re.kr,
{franck.le-gall, mengxuan.zhao}@eglobalmark.com

*Abstract* — Global IoT services (GIoTS) are combining locally available IoT resources with Cloud-based services. They are targeting world-wide services. GIoTS require interoperability between the locally installed heterogeneous IoT systems. Semantic processing is an important technology to enable data mediation as well as knowledge-based processing. This paper explains a system architecture for achieving world-wide semantic interoperability using international standards like oneM2M and the OMA NGSI-9/10 context interfaces (as used in the European Future Internet Platform FIWARE). Semantics also enables the use of Knowledge-based Semantic Processing Agents. Furthermore, we explain how semantic verification enables the testing of such complex systems.

## I. INTRODUCTION

The Internet of Things (IoT) is considered to be the next step in the Internet evolution. A common complaint is the lack of accepted standards and the huge fragmentation of the IoT market. Many international organizations are working on defining standards for the IoT. Notably, oneM2M [1] is defining an International Standard for IoT data exchange on a world-wide scale. The second problem tackled by oneM2M is integration of existing IoT platforms (e.g. OIC and AllJoyn) as well as the utilization of existing standards (e.g. MQTT, CoAP and Lightweight M2M). As oneM2M focuses on the communication aspect of IoT, the complete area of providing interoperable data services on and between IoT clouds is not covered yet. Here the large scale European Future Internet Platform FIWARE is offering a set of well-aligned Cloud enablers that can be used for receiving, processing, contextualizing and publishing IoT data. FIWARE's large set of enablers play together following the agreed OMA NGSI-9/10 standard. Different to oneM2M, OMA NGSI enables content and type-based queries. Europe and Korea have just started a joint project which is tackling the "World-Wide Interoperability for Semantic IoT" (WISE-IoT) using oneM2M and FIWARE.

Since the initial design of the FIWARE platform five years ago, IoT has moved from efficiently transporting sensor data into the Cloud to a distributed system that is putting the data to use, e.g. by providing analytic functions, linking various data sources, as well as contextualizing the information. A key technology for such smart processing is semantic representation as well as semantic reasoning. This is typically done with a Triple-Store and respective tools for semantic processing. The long-term target is to make the information of the IoT accessible to automatic knowledge processing agents which can assist users and recommend needed actions. Fig. 1 gives an overview about abstraction layers in a world-wide IoT system.

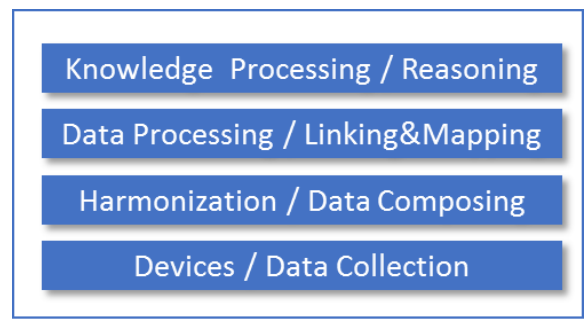

**Figure 1: IoT Abstraction Layers**

This discussion raises the issue of how we can achieve seamless interworking and what technical components are missing. It is the aim of this paper to explain the recent advancements in standardization, research as well as field testing. We will discuss this in the context of so-called

"Global IoT Services (GIoTS)". GIoTS work on the premise that in the future IoT service will use IoT resources available around the world while essential functions of the service will be executed in Cloud data centers. GIoTS are business concepts that enable companies to offer fully digitalized IoT services around the world, while keeping essential trade secrets (e.g. analytics, optimization routines) confidential stored in their data centers. GIoTS need to deal with (a) potential disruptions on the international communication lines, (b) with dynamically utilizing locally available IoT resources, and (c) with constant adaptation of the provided services by improving the analytics as well as the actuation functions. Customer IoT services will benefit from semantic interoperability by seamlessly using available IoT resources independent of the SDK or technolgoy used to implement and connect the devices. Further advantages can be realized through more sophisticated search functions as well as automated mashup.

The paper is structured as follows. We describe related state-of-the-art in the following section 2. Section 3 explains the technical approach for GIoTS as well as for interworking between oneM2M, FIWARE and semantic knowledge processing. Section 4 explains the system architecture. Section 5 explains important details of the realizations like edge gateways with semantic mediation functions, semantic validation, as well as experiments and results. Section 6 will conclude the paper with a summary and an outlook.

## II. RELATED STANDARDS AND IOT PLATFORMS

*oneM2M*

oneM2M is a cooperation between different standard development organizations (SDOs) in the world to define a world-wide standard for M2M/IoT communication. The founding SDO members include ATIS, TIA (USA), ETSI (Europe), CCSA (China), ARIB, TTC (Japan), and TTA (Korea). In 2015, TSDSI (India) has newly joined oneM2M. It has also a partnership with five industrial consortia, including the BroadBand Forum (BBF), the Continua Health Alliance, Home Gateway Initiative (HGI), the New Generation M2M Consortium, and the Open Mobile Alliance (OMA). Accordingly, oneM2M can jointly develop technical specifications for horizontal M2M/IoT service platforms across a number of industry verticals.

The oneM2M standards adopted a hierarchical resource structure, as shown in Fig. 2. In the figure, CSE (common service entity) represents an instantiation of a set of common service functions (CSFs) of M2M/IoT systems, e.g., device registration, discovery, and management; AE represents an entity in the application layer residing in a number of nodes and providing service logics; group represents a group of oneM2M resources of the same or mixed types; container represents a container for data instances and contentInstance represents a data instance in the container [3]. The oneM2M resources can be uniquely accessed by the uniform resource identifier (URI) along with the basic five CRUDN (Create, Retrieve, Update, Delete, Notify) operations [11].

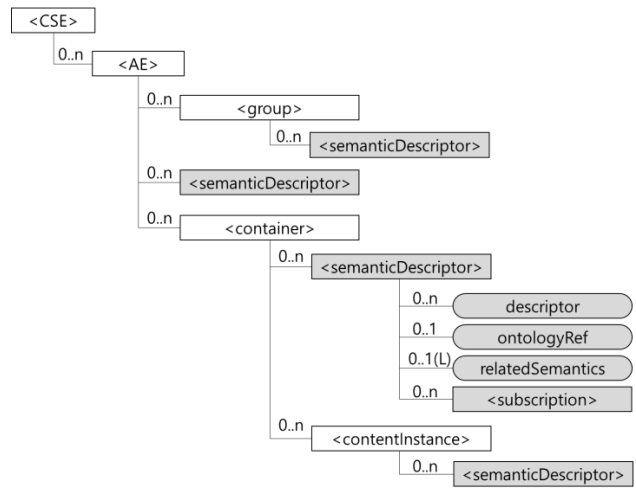

**Figure 2: Resource Tree with Semantic Descriptors**

Release 1 of oneM2M [3] published in Jan. 2015, provides very limited options of describing resources and the data or services they provide. Primarily such descriptions are restricted to a few pre-defined specific attributes like the name and ID of resource, and annotation with simple string labels. This limits the functionalities to discover resources to simple filters based on the specific attributes and the matching of string labels. The result is a close coupling of applications that communicate via oneM2M resources. Those applications need to have a-priori know-how about the resource structures and the data format used.

While this may be suitable for more traditional vertical M2M setups, this is problematic for a true Internet of Things scenario (and especially GIoTS) with different applications reusing the same information. Especially large-scale deployment with changing information sources need better features to understand the provided IoT data. As an example, due to the heterogeneity of the system, different types of resources may provide the same kind of information using different data formats. Consuming applications need to be able to find all possible relevant sources and utilize them dynamically.

**Semantic Annotations**: Rel. 2 of oneM2M added the concept of semantically annotating relevant oneM2M resources [9]. Semantic annotations are stored in *semantic descriptor* child resources of the annotated resource (see Fig. 2). A semantic descriptor consists of semantic subject-predicate-object triples represented in RDF [7]. They provide the flexibility to describe all relevant aspects of the resource, for example which real-world object the resource is related to, how the resource can be used for interacting with a real-world object, and many more. Ontologies are used as a common vocabulary. They define the relevant concepts and properties, e.g. the type of the real-world object, relations between objects, the unit of measure etc.

**3[rd] Party Annotations**: It is worth noting that semantic descriptors can be added by third-party applications, i.e. a semantic annotation tool. Resources of existing legacy applications can be augmented by 3rd parties thus enabling their use by semantic applications even when such use was not planned for at deployment time.

**Semantic Discovery**: oneM2M Release 2 defines discovery criteria (so-called filters) which are described in the semantic query language SPARQL [8]. Discovery in oneM2M operates locally and starts with the children of a specified resource in the resource structure, possibly the <CSE> root resource. Remote resources may be found through their announced counterparts available locally. For semantic discovery, the SPARQL filter is applied to the semantic description in RDF stored in the semantic descriptor child resource of the targeted resource and if the SPARQL finds a match, the semantically described parent resource is added to the result set. This greatly extends the expressiveness of the oneM2M discovery feature [10].

**Summary**: oneM2M provides the base for a world-wide interoperable IoT system using international standards. The initial design concept of having only black-box data container has been extended to enable semantic annotations. This opens the way of using the rich semantic description for improving the handling of IoT data, especially for contextualization as well as for verification and testing. However, the query, filter and subscribe operations on oneM2M data are not working on the data values, but only on the meta-data itself, thus limiting the efficiency of the respective operations.

*Future Internet Core Platform FIWARE*

FIWARE is the result of the Future Internet PPP, a private-public partnership to create a novel Cloud platform for future Internet business in Europe. In the meantime, the FIWARE platform is being used around the world, from Mexico, to Senegal and New Zealand. FIWARE is built on the concept of configuring a service platform from a large library of existing generic enablers. Those generic enablers are built with the intention to easily work together and re-use common functionalities. After the initial development and capacity building phase, FIWARE has now entered the acceleration phase in which many new businesses are on-boarded on the platform. Companies have started to build commercial offers using FIWARE. For example, NEC is offering its *Cloud City Operation Center (CCOC)*. The CCOC is offering Smart City services such as a City Dashboard, Smart Waste Management, Traffic Monitoring and Prediction, as well as Smart Street Light. The CCOC is using the IoT Broker generic enabler to retrieve IoT information from cities like Santander (Spain) or Wellington (New Zealand). Further enablers are used to process the received information, e.g. by providing big data analysis on historical data.

Compared to oneM2M, FIWARE is focusing on providing the data cloud for processing incoming data. Big data storage, visualization components, as well as components to publish data as Open Linked Data enhance the usability of data. Important advantages of FIWARE are its plug&play architecture, the use of contextualized information models, as well as rich query and subscriptions. Contextualized information contains the raw value of the information, but also context information like location or meta-data about accuracy or reporitng entities. The data and context chapter is providing an enabler to process structured and unstructured data. The ORION context broker is a simple Cloud component that aggregates contextualized data in a big data store and provides access to the data with the OMA NGSI-10 API [2]. The NEC IoT Broker implements the full OMA NGSI concept including NGSI-9 for registering and discovering context entities. It provides a scale-out implementation in which stateless instances of the IoT Broker can be instantiated to handle increasing load. Furthermore, it can serve as a Federation Agent that can execute distributed queries including many different NGSI-enabled components. The IoT Broker utilizes a specific context information model, which enables expressive functions based on this model. Table 1 compares it with existing message brokers.

FIWARE uses the OMA Next Generation Service Interface (NGSI-9/10) data model to build the common information

| System | Function Description | Description | Comment |
| --- | --- | --- | --- |
| **ROS (Robot operating System)** | topic-based pub/sub system, no discovery | Enables async. operation of various robot systems | Droones, autonomous driving cars, … |
| **MQTT** | Pub/Sub system, no discovery | Highly optimized for telemetric data | Integration w. oneM2M |
| **Rabbit MQ** | Queue and topic-based pub/sub broker, no filtering, no discovery | Broker with lots of language support, cluster management | Main prupose Cloud services |
| **Kafka/Zookeeper** | message broker & cluster mangement, no filtering, no discovery | Focus on performance, limited functionality, reliable delivery of messages | Mostly used in Cloud-based system |
| **IoT Broker** | pub/sub using NGSI format , discovery: scope /restrictions filters | Focus on context information, federation, geolocation, geodiscovery, geofencing, fire&forget event delivery | IoT chapter of FIWARE platform, can be combined with a storage agent |

**Table 1: Comparing different Message Brokers**

model of IoT-based systems. In a Smart City, FIWARE exposes the City Model in NGSI format. Apps can understand and actuate systems in the city by manipulating the City Model. Wise-IoT is creating modern installations of FIWARE which are using a semantically grounded information model. NGSI entity types, attribute types, as well as metadata types are defined by ontologies managed by a separate *IoT Knowledge server*. Such a semantic grounding enables to use of relationships included in ontologies during the operation of the FIWARE system. Examples of exploring semantic relationships is the use of a subtyping hierarchy between entity types when searching for IoT entities.

**Summary**: FIWARE has taken a different approach then oneM2M by focusing on a common data model and powerful interfaces for searching and finding information. A rich set of enablers following the same APIs and data model creates a platform for developing IoT applications.

FIWARE lacks the ability of oneM2M for a world-wide communication standard. FIWARE needs to be extended to support semantic data models.

*Knowledge-Based Semantic Processing Agents*

Semantic grounding is not enough to create truly smart services. Such services reason on the provided input and combine real-time IoT information with other data sources (e.g. from sentiment analysis of social networks or data from mobile services). In order to integrate many different heterogeneous systems, semantic reasoning as well as knowledge processing is needed for correctly mapping different data structures, as well as to provide semantic rich queries. Therefore, we need to extend FIWARE with additional Knowledge-Based Semantic Processing Agents (KSPA). Those agents receive contextualized information (provided by FIWARE). Using rich processing frameworks for semantic inferences as well as knowledge reasoning, KSPA provide domain-specific smart processing. The results of KSPA computation can be fed back into the FIWARE world model and made available to a large set of applications. In special cases, KSPA can expose a communication endpoint where by semantic services or queries using semantic query languages like SPARQL can be exposed.

**Summary**: The concept of KSPA extends the oneM2M and FIWARE concept with the ability of processing semantic informtion and reason on it. KSPA can either work as hidden processing agents or expose their services to applications.

*Previous Work by the Authors*

In this paper, the authors present the interoperability architecture for oneM2M, FIWARE and KSPA agents. Compared to [4], [5], our current paper introduces and discuss in greater length the use of international standards for IoT and elaborate especially the use of semantic annotation. Furthermore, we explain the concept of Semantic Mediation Gateways (SMG) as well as the use of semantic verification. In our keynote at the ICSC'2016 conference [6], we gave a high-level introduction into semantic processing in oneM2M and FIWARE. We did not explain the concept of SMG and did not explain in details the use of semantics in oneM2M, FIWARE and KSPA.

### III. TECHNICAL APPROACH

We are targeting Global IoT Services (GIoTs) characterized by the following requirements:

**World-wide Use**: GIoTS service can be requested anywhere in the world. They need to cope with the locally available resources as well as communicate back to Cloud-based service elements.

**Re-Use of existing IoT resources**: IoT resources are offered using many different protocols and eco-systems. We need to re-use the available resources.

**Discovery & Dynamic Composition**: GIoTS need to discover locally available IoT resources and compose the IoT service dynamically. This is especially true for edge computing devices as well as for mobile apps.

**Semantic Interoperability**: GIoTS dynamically discover the local environments and need to cope with previously unknown IoT devices and data sources. It is highly important to ensure that the use of these devices is well understood and that they are correctly used. Semantic definition of their type, attributes, as well as metadata is needed to automatically integrate them as well as to ensure compatibility

**Recommendation and DevOps**: In many cases, automatic adaptation to the local IoT resource situation is not feasible. Manual support of either the end-users or developers is needed. Semi-automatic support functions such as recommendations can help to compose a GIoT. Whenever automatic adaptation is not possible, we need human developers to help in adapting the GIoTS to the local context. Such features are today covered under the term "DevOp".

**Adaptive Communication and Disconnected Operation:** GIoTS components need to use the available network and adapt their communication behavior as needed. They need to deal with temporary loss of communication (Disconnection) as well as with low available bandwidth as well as with long latencies.

The technical approach to fulfil these requirements is as follows: (a) we solve the **world-wide communication** requirements by using oneM2M. As explained, oneM2M provides world-wide connectivity, security, as well as an agreed model for transporting IoT data. (b) oneM2M is furthermore already standardizing **syntactic interoperability** between many different IoT standards. To solve the semantic interoperability problem, oneM2M is starting the standardization of **semantic annotations** (Rel.2) and **semantic operation** (e.g. mashup, analytics, semantic queries) (Rel.3). To the best of our knowledge, the work described in this paper is the first time that oneM2M semantic annotations have been implemented and used for semantic mediation between different systems, (c) we solve the **dynamic discovery and composition** problem by re-using the respective features provided by FIWARE. The extensions done to today's FIWARE system is to base FIWARE's type and metadata system on concepts from semantic specifications ("*Explicit Semantic Grounding*"). FIWARE works on the context level modelling the world as (virtual) entities. The relative simple FIWARE data model combined with semantic grounding providing the right abstractions to deal with large sets of unknown entities. The problem to solve is the semantic mediation between oneM2M and FIWARE. A *Semantic Mediation Gateway (SMG)* was shown at the ETSI M2M workshop 2015, but was not build on the concept of a programmable edge gateway.

(d) FIWARE itself does not provide the capabilities for **knowledge processing** and **logical inferences**. We therefore provide those capabilities through the use of KSPA agents. (e) Finally, in order to deal with **disconnected operation**, bandwidth and latency requirements, we use the concept of **dynamic programmable edge gateways**. Those gateways can host dynamically deployed components using virtualization techniques like Docker. One example of such a component can be a mediation function. Self-contained components can deal with disconnections. Smart compression functions can reduce bandwidth consumption. Local control loops can operate with low latency. Through the use of dynamic deployment, we can also provide DevOps capabilities.

## IV. SYSTEM ARCHITECTURE

The following picture explains the system architecture. At the bottom, in the *Device Layer*, we have a heterogeneous set of devices (sensors, actuators, edge gaeteways). They are connected with different networking technologies to the *Homogenization Layer* that provides the capabilities for integrating various data formats to harmonize the access capabilities (API, Access Control, Real-Time Streams) as well as to provide connectivity on a large scale. The *Selection and Aggregation Layer* selects and aggregates the information required by the *Analytics and Knowledge Layer.* For this purpose, it has an underlying information model and an API that allows specifying the required information according to this model. The API may provide further functionality for scoping the query, i.e., a-priori restricting the search space with respect to retrieving the information from the Homogenization Layer, and filtering, i.e., filtering the information after it was retrieved. In order to make information from the Homogenization Layer accessible to the Selection and Aggregation Layer, *Semantic Mediation Functions* deployed on edge or cloud gateways are used. Finally, the *Analytics and Knowledge Layer* provides analytics functions that are applied to the selected information retrieved through the Selection and Aggregation Layer. For example, the Analytics and Knowledge Layer can have triple-stores and utilize knowledge-based reasoning to derive new know-how based on the selected information.

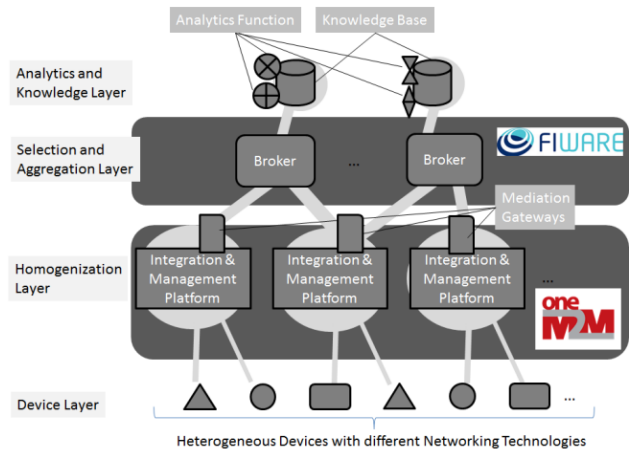

Figure 3: System Architecture

In our work, the Homogenization Layer is realized using a oneM2M system. As described above, oneM2M enables the integration of a large number of different device and platform technologies, exposing them through its Mca interface. The Mca is a REST-API supporting a number of different bindings, e.g. HTTP, CoAP and MQTT. The devices and the information they provide are represented as REST resources, where data is typically stored in containers, as black-box contentInstances containing the actual data, e.g. the measured value of a sensor. With Release 2 of oneM2M, the resources can be semantically annotated.

The Selection and Aggregation Layer is realized using FIWARE general enablers based on the context interfaces defined as part of OMA NGSI. The needed bindings were developed in FIWARE. Depending on the requirements and scale, either the Orion Context Broker GE or the IoT Broker GE can be chosen as a basis. While the former has a centralized architecture, the latter has a distributed underlying model that allows creating a federated structure, e.g. corresponding to a hierarchical organization where each level is in charge of its data. The NGSI information model is based on an entity – attribute model, where the entities are used to model real world objects as well as virtual entities and the attributes describe different aspects of the entity. Entities have types that can be defined as ontology concepts, enabling the creation of a type hierarchy using the subtype relation of ontologies. For example, the entity type may be *meeting room*, that has an attribute *occupancy*. A *meeting room* is subtype of the more general type *room* that introduces the attribute *room temperature*, which is then inherited by meeting room. OMA NGSI supports queries for entities based on identifiers and types, e.g. give me the *room temperature* of room123 (entity identifier) or give me the *occupancy* of *meeting rooms*. To reasonably limit the result set of the latter query, NGSI supports the concept of scopes. A typical example of scopes is a location scope that would, e.g., enable limiting the scope to the *meeting room* entities on a specific floor in a building.

In order to enable the required interaction between FIWARE and oneM2M, Semantic Mediation Gateways are used, which are described in detail in the next section.

## V. REALIZATION ASPECTS OF SELECTED COMPONENTS

The following section describe the realization of selected components of the architecture:

### Semantic Mediation Gateways (SMGs)

Mediation functions transform data from the representation used in one system into the representation used in another system. The special feature of a Semantic Mediation Gateway (SMG) is that this transformation is guided by semantic annotation as well as knowledge built into the gateway. This is used to make information from the oneM2M system available to FIWARE, The entity-attribute based information structure is created by semantically annotating the oneM2M resources in such a way that the required information like entity id, entity type, attribute name and required metadata can be infered by the SMG using suitable reasoning.

An SMG, as shown in Figure 4, consist of: (a) a **discovery** component that analyzes the system where data originates from (Originator), (b) a selector that identifies which transformation process to apply, (c) a library of **transformation processes**, as well as of data conversion routines, (d) a **semantic reasoner** equipped with a knowledge base, (e) and finally an **updater** that updates the information in the target system.

(1) The discovery component either uses subscription features of the Integration & Management Platform or browses the available data. In the case of oneM2M, the discovery component searches for semantically annotated resources representing information sources like sensors. (2) Once discovered the semantic information is used to find an appropriate transformation process with the data conversion routines needed for the data items – if available. (3) The selected transformation process is then instantiated subscribing for new information items from the Integration & Management Platform. In the case of oneM2M the subscription is for new contentInstance resources, which contain the actual values, e.g. measured by a sensor.

(4) Whenever a new or changed data item is discovered, the item is retrieved and forwarded to the instantiated transformation processes. (5) Based on the data item and the semantic description, the transformation processes uses the semantic reasoner to derive the information needed for the target information model. This especially infers the name of the target entity, its entity type, the target attribute, as well as the meta-data that can be created. Data conversion routines are then used to transform the data into the data representation of the target system. (6) In the case of oneM2M, the semantic annotation is used to derive the elements needed for the NGSI data model. The attribute value and some of the metadata values are then extracted from the retrieved contentInstance and converted into the NGSI data representation, which is then forwarded to the updater. (7) The updater provides the information to the Broker in the Selection and Aggregation Layer using the appropriate protocol, e.g. HTTP. Two modes are supported: (a) pull requests from the Broker, e.g. using NGSI queries, or (b) push operations initiated by the SMG, e.g. using NGSI updates.

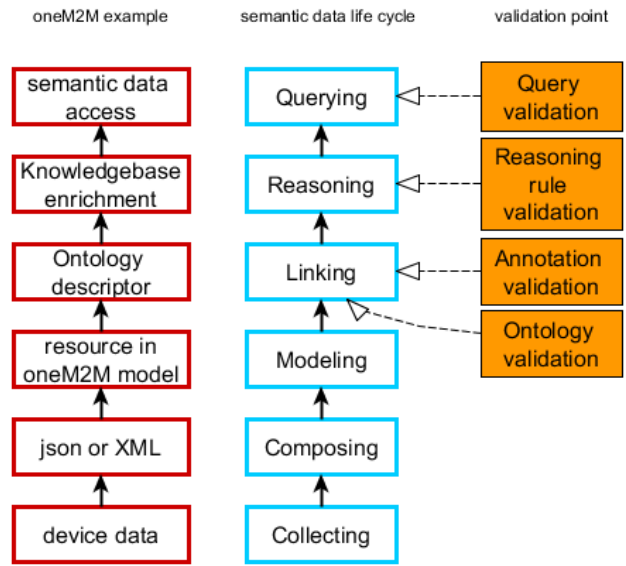

**Figure 5: Semantic Validation Lifecycle**

### Semantic Validation

The use of semantic technologies is not only enabling knowledge processing and semantic interoperability. It also allows testing and verifying a complex system like GIoTS by the use of semantic verification. The first step is to ensure the semantic information itself and the use of such information respect common-recognized rules. In the Fiesta-IoT project, we identify several points of validation within the semantic data lifecycle - from data production to consummation (see Figure 5). The validation points are the following: 1) If a specific concept is needed to be expressed by extending a reference ontology (i.e. oneM2M base ontology), the ontology should be first checked whether contradictory terms and relations exist; 2) data annotated with the validated ontology should be checked if all used semantic terms are from the declared ontologies and if no errors are present (i.e. incompatible data type against the declared data type); 3) With assumption of an open "rule base" in which new rules can be added, rules should be checked that they do not generate contradictory results compared to current rules in the base; 4) Before queries are forwarded to the semantic database to retrieve information, they should be checked against the SPARQL syntax in order to not generate unexpected consequences to the database. Following these steps, a minimal level of semantic interoperability can be guaranteed in semantic supporting systems.

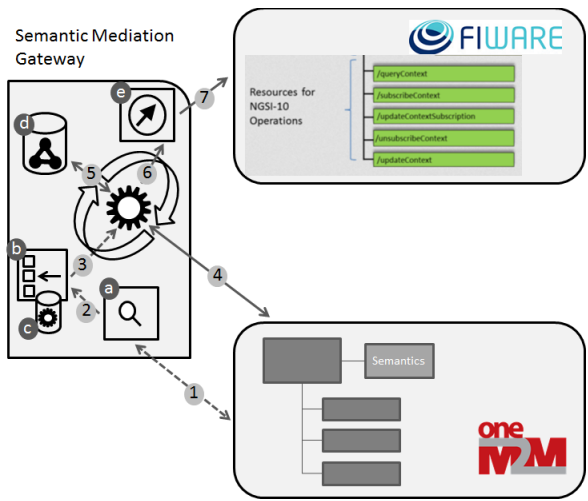

**Figure 4: Semantic Mediation**

Initial work has been done to develop the validation tool chain and Easy Global Market is now providing test tools for base line ontologies to validate linked-data descriptions and ontologies against reference ontologies in respect with logical and lexical validation aspects. This tool is provided as a webservice that a user can simply submit the semantic file to, either through a built-in web page or through any HTTP client. A validation report is returned to the user containing the total processing duration, the details of errors (if they exist) categorized by their nature (syntactic, semantic, etc).

*Experiments and Results*

The system as described has been realized using the open source implementation of oneM2M called Moebius. We used the FIWARE Context Broker ORION for handling the aggregation of data and the Wirecloud toolset for visualisation. KSPA agents have been realized using the runtime environment from FIESTA_IoT. During the ETSI M2M workshop in December 2015, we demonstrated the SMG and also showed how a mobile phone-based tool can be used to annotate oneM2M resources. After annotation, the data became immediately available in the FIWARE system and accessible to KSPA agents.

## VI. DISCUSSION AND CONCLUSIONS

In this paper we described a world-wide interoperable system for the Internet-of-Things based on international standards. oneM2M was used as the data harmonization layer and the FIWARE system as the contextualization layer where data can be aggregated and exposed to applications. Knowledge-based Semantic Processing Agents can further process the data. The concept of the Semantic Mediation Gateways helps in intelligently mediating data from one system to the other. We have furthermore shown that beyond the semantic interoperability, semantic verification can help in testing the system and ensure correct operation. A first prototype of the system was shown at the ETSI M2M workshop in Dec 2015 and is the base of a comprehensive collaborative project between Korea and Europe called WISE-IoT starting in summer 2016.

## ACKNOWLEDGEMENTS

Mr Kim was supported by Institute for Information & communications Technology Promotion (IITP) grant funded by the Korea government (MSIP) (No.B0184-15-1003).

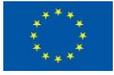 This work has also received funding from the European Union's Horizon FP7 research program within the project FI-CORE under grant agreement No 632893 as well as H2020 research program within the project FIESTA under grant agreement No 643943. We acknowledge using material from the not yet started project WISE-IoT. The content of this paper does not reflect the official opinion of the European Union. Responsibility for the information and views expressed in the therein lies entirely with the author(s).

## BIOGRAPHIES

JAEHO KIM [M] (jhkim@keti.re.kr) is a managerial researcher and a team leader in IoT Platform Research Center at the Korea Electronics Technology Institute. He is now serving as IoT Convergence Service Project Group chair of TTA and Device Working Group chair of Korea IoT Association. He received the BS and MS degrees from the Hankuk University of Foreign Studies. Currently, he is a Ph.D. candidate in the electrical & electronic engineering from the Yonsei University.

Jaeseok Yun [M] (jaeseok@keti.re.kr) is a senior researcher in the IoT Platform Research Center at Korea Electronics Technology Institute (KETI). Prior to his current position, he worked as a postdoctoral research scientist in the Ubiquitous Computing Research Group in the School of Interactive Computing at Georgia Institute of Technology, GA, USA. He earned his M.S. and Ph.D. in Mechatronics from Gwangju Institute of Science and Technology (GIST).

MARTIN BAUER [M] (martin.bauer@neclab.eu) is a Senior Researcher focused on IoT and context management related research and standardization at the NEC Laboratories Europe in Heidelberg, Germany. He participates in oneM2M standardization focusing on aspects related to semantics. Prior to his current position he worked as a researcher at the University of Stuttgart, from where he also received his PhD. Furthermore, He holds M.S. degrees from both the University of Oregon and the University of Stuttgart.

ERNÖ KOVACS [M] (ernoe.kovacs@neclab.eu) is Senior Manager at NEC Laboratories Europe, Heidelberg, Germany. He is leading the research group "Cloud Services and Smart Things". His team is involved in oneM2M standardization, the European Future Internet Core Platform FIWARE, as well as in many international research projects. He received his PhD degree from the University of Stuttgart. Before joining NEC, he was working for Sony' research lab in Stuttgart. He holds 27 granted patents.

FRANCK LE GALL [M] (franck.le-gall@eglobalmark.com) is COO at Easy Global Market where he is driving company development of advanced testing technologies as well as integration of IoT and data platforms. He is involved in standardisation including oneM2M and FIWARE. He holds a PhD degree from university of Rennes. He has participated in large cooperative R&D projects and directed more than 10 large scale projects. He studied evaluation of innovation, technical programs and research projects.

MENGXUAN ZHAO (mengxuan.zhao@eglobalmark.com) has a PhD in computer science from University of Grenoble. Her thesis "Discrete control for the Internet of things and smart environment via a shared infrastructure" brought the classical discrete control theory to a new application domain within IoT using semantic techniques. With a strong background in the IoT domain and semantic technologies, she joined Easy Global Market after her thesis preparation in Orange Labs, Grenoble. She participats in standardization.